\documentstyle[prd,aps,preprint,tighten,epsfig]{revtex}

\begin{document}

\draft

\title{Correlation between the Charged Current Interactions \\ of Light and
Heavy Majorana Neutrinos}
\author{{\bf Zhi-zhong Xing}
\thanks{E-mail: xingzz@ihep.ac.cn}}
\address{\sl Institute of High Energy Physics, Chinese Academy of
Sciences, Beijing, China} \maketitle

\begin{abstract}
The evidence for neutrino oscillations implies that three neutrino
flavors ($\nu^{}_e, \nu^{}_\mu, \nu^{}_\tau$) must have different
mass states ($\nu^{}_1, \nu^{}_2, \nu^{}_3$). The most popular
idea of generating tiny masses of $\nu^{}_i$ is to introduce three
heavy Majorana neutrinos $N^{}_i$ (for $i = 1, 2, 3$) into the
standard model and implement the seesaw mechanism. In this
approach the neutrino mixing matrix $V$ appearing in the charged
current interactions of $\nu^{}_i$ is not unitary, and the
strength of unitarity violation of $V$ is associated with the
matrix $R$ which describes the strength of charged current
interactions of $N^{}_i$. We present an explicit parametrization
of the correlation between $V$ and $R$ in terms of nine rotation
angles and nine phase angles, which can be measured or constrained
in the precision neutrino oscillation experiments and by exploring
possible signatures of $N^{}_i$ at the LHC and ILC. Two special
but viable scenarios, the Type-I seesaw model with two heavy
Majorana neutrinos and the Type-II seesaw model with one heavy
Majorana neutrino and one Higgs triplet, are taken into account to
illustrate the simplified $V$-$R$ correlation. The implications of
$R\neq 0$ on the low-energy neutrino phenomenology are also
discussed. In particular, we demonstrate that the non-unitarity of
$V$ is possible to give rise to an appreciable CP-violating
asymmetry between $\nu^{}_\mu \rightarrow \nu^{}_\tau$ and
$\overline{\nu}^{}_\mu \rightarrow \overline{\nu}^{}_\tau$
oscillations with short or medium baselines.
\end{abstract}


\newpage

\framebox{\large\bf 1} ~ Very robust evidence for non-zero
neutrino masses and large lepton flavor mixing has recently been
achieved from solar \cite{SNO}, atmospheric \cite{SK}, reactor
\cite{KM} and accelerator \cite{K2K} neutrino oscillation
experiments. This great breakthrough opens a new window to physics
beyond the standard model (SM), because the SM itself only
contains three massless neutrinos whose flavor states
$\nu^{}_\alpha$ (for $\alpha = e, \mu, \tau$) and mass states
$\nu^{}_i$ (for $i=1, 2, 3$) are identical. The most natural and
popular way to generate non-vanishing but tiny masses $m^{}_i$ for
$\nu^{}_i$ is to extend the SM by introducing three right-handed
neutrinos but preserving its $SU(2)_{\rm L} \times U(1)_{\rm Y}$
gauge symmetry. This class of neutrino mass models is commonly
referred to as the Type-I seesaw models \cite{SS1}. If one Higgs
triplet $\Delta^{}_{\rm L}$ and three right-handed neutrinos are
simultaneously introduced into the SM, the resultant $SU(2)_{\rm
L} \times U(1)_{\rm Y}$ gauge models are usually classified as the
Type-II seesaw models \cite{SS2}. In either case the mass states
of three right-handed neutrinos, denoted as $N^{}_i$ (for
$i=1,2,3$), have the positive eigenvalues $M^{}_i$ which can be
much larger than the electroweak symmetry breaking scale $v
\approx 174$ GeV. The smallness of $m^{}_i$ is therefore
attributed to the smallness of $v^2/M^{}_i$ (and the smallness of
the vacuum expectation value of $\Delta^{}_{\rm L}$ in the Type-II
seesaw models). Both light and heavy neutrinos are the Majorana
particles in such seesaw models, in which the lepton number ($L$)
is not conserved. Associated with the seesaw mechanism, the
leptogenesis mechanism \cite{FY} may naturally work to account for
the cosmological matter-antimatter asymmetry via the CP-violating
and out-of-equilibrium decays of $N^{}_i$ and the
$(B-L)$-conserving sphaleron processes \cite{Kuzmin}.

To directly test the seesaw and leptogenesis mechanisms, it is
desirable to experimentally discover the heavy Majorana neutrinos
$N^{}_i$ at high-energy $e^+e^-$ and (or) hadron colliders. Since
such neutral and weakly interacting particles leave no trace in
ordinary detectors, their possible collider signatures must
involve charged leptons via the charged current interactions for
some $\Delta L = 2$ processes (i.e., those processes with the
lepton number being violated by two units) \cite{Han}. In the
basis of mass states, the standard charged current interactions of
$\nu^{}_i$ and $N^{}_i$ can be written as
\begin{eqnarray}
-{\cal L}^{}_{\rm cc} \; = \; \frac{g}{\sqrt{2}} \left[
\overline{\left(e~~ \mu~~ \tau\right)^{}_{\rm L}} ~V \gamma^\mu
\left( \matrix{\nu^{}_1 \cr \nu^{}_2 \cr \nu^{}_3} \right)^{}_{\rm
L} W^-_{\mu} + \overline{\left(e~~ \mu~~ \tau\right)^{}_{\rm L}}
~R \gamma^\mu \left( \matrix{N^{}_1 \cr N^{}_2 \cr N^{}_3}
\right)^{}_{\rm L} W^-_\mu \right] ~ + ~ {\rm h.c.} \; ,
\end{eqnarray}
where $V$ is just the Maki-Nakagawa-Sakata (MNS) neutrino mixing
matrix \cite{MNS} responsible for neutrino oscillations, and $R$
describes the strength of charged current interactions between
$(e, \mu, \tau)$ and $(N^{}_1, N^{}_2, N^{}_3)$. It has been
noticed that $V$ is not exactly unitary in the seesaw models and
its deviation from unitarity is essentially characterized by
non-vanishing $R$ \cite{Zhou}. Indeed, the production and
detection of $N^{}_i$ at the Large Hadron Collider (LHC) or the
International Linear Collider (ILC) require not only $M^{}_i
\lesssim {\cal O}(10)$ TeV but also appreciable sizes of the
matrix elements of $R$. Hence the unitarity violation of $V$ might
show up in the future long-baseline neutrino oscillation
experiments, if the collider signatures of heavy Majorana
neutrinos are really accessible. In this sense, the correlation
between $V$ and $R$ signifies an important relationship between
neutrino physics and collider physics.

The purpose of this work is just to reveal how the charged current
interactions of light and heavy Majorana neutrinos are correlated
with each other. We shall show that the correlation between $V$
and $R$ can in general be parametrized in terms of nine rotation
angles and nine phase angles. This parametrization is independent
of any details of a Type-I or Type-II seesaw model, and its
parameters can be measured or constrained in the precision
neutrino oscillation experiments and by exploring possible
signatures of $N^{}_i$ at the LHC and ILC. We shall take two
special but viable examples, the Type-I seesaw model with two
heavy Majorana neutrinos and the Type-II seesaw model with one
heavy Majorana neutrino and one Higgs triplet, to illustrate the
simplified form of $V$-$R$ correlation. The implications of $R\neq
0$ on the low-energy neutrino phenomenology will also be
discussed. In particular, we shall demonstrate that the
non-unitarity of $V$ is possible to give rise to an appreciable
CP-violating asymmetry between $\nu^{}_\mu \rightarrow
\nu^{}_\tau$ and $\overline{\nu}^{}_\mu \rightarrow
\overline{\nu}^{}_\tau$ oscillations with short or medium
baselines.

\vspace{0.4cm}

\framebox{\large\bf 2} ~ Without loss of generality, we choose the
basis in which the flavor and mass states of three charged leptons
are identical. In this basis, the neutrino mass terms generated
from spontaneous $SU(2)^{}_{\rm L} \times U(1)^{}_{\rm Y}
\rightarrow U(1)^{}_{\rm em}$ symmetry breaking can be written as
\begin{eqnarray}
-{\cal L}_{\rm mass} \; = \; \frac{1}{2} ~ \overline{\left(
\nu^{}_{\rm L} ~N^c_{\rm R}\right)} ~ \left( \matrix{ M^{}_{\rm L}
& M^{}_{\rm D} \cr M^T_{\rm D} & M^{}_{\rm R}}\right) \left(
\matrix{ \nu^c_{\rm L} \cr N^{}_{\rm R}}\right) ~ + ~ {\rm h.c.}
\; ,
\end{eqnarray}
where $\nu^c_{\rm L}$ and $N^c_{\rm R}$ are defined as $\nu^c_{\rm
L} \equiv C \overline{\nu^{}_{\rm L}}^T$ and $N^c_{\rm R} \equiv C
\overline{N^{}_{\rm R}}^T$ with $C$ being the charge conjugation
matrix, $M^{}_{\rm D} = Y^{}_\nu v$ and $M^{}_{\rm L} =
Y^{}_\Delta v^{}_{\rm L}$ result from the Yukawa interactions of
Higgs doublet and triplet with $v\approx 174 ~{\rm GeV}$ and
$v^{}_{\rm L} \lesssim 1 ~ {\rm GeV}$ being the corresponding
vacuum expectation values, and $M^{}_{\rm R}$ is the mass matrix
of three right-handed Majorana neutrinos. The overall $6\times 6$
neutrino mass matrix in ${\cal L}^{}_{\rm mass}$, denoted as
${\cal M}$, is symmetric and can be diagonalized by a unitary
transformation ${\cal U}^\dagger {\cal M} {\cal U}^* =
\widehat{\cal M}$; or explicitly,
\begin{eqnarray}
\left(\matrix{V & R \cr S & U}\right)^\dagger \left( \matrix{
M^{}_{\rm L} & M^{}_{\rm D} \cr M^T_{\rm D} & M^{}_{\rm R}}\right)
\left(\matrix{V & R \cr S & U}\right)^*  = \left( \matrix{
\widehat{M}^{}_\nu & {\bf 0} \cr {\bf 0} & \widehat{M}^{}_{\rm
N}}\right) \; ,
\end{eqnarray}
where $\widehat{M}^{}_\nu = {\rm Diag}\{m^{}_1, m^{}_2, m^{}_3\}$
and $\widehat{M}^{}_{\rm N} = {\rm Diag}\{M^{}_1, M^{}_2,
M^{}_3\}$ with $m^{}_i$ and $M^{}_i$ (for $i=1, 2, 3$) being the
light and heavy Majorana neutrino masses, respectively. After this
diagonalization, one may express the neutrino flavor states
$\nu^{}_\alpha$ (for $\alpha = e, \mu, \tau$) in terms of the
light and heavy neutrino mass states $\nu^{}_i$ and $N^{}_i$ (for
$i=1, 2, 3$):
\begin{eqnarray}
\left(\matrix{ \nu^{}_e \cr \nu^{}_\mu \cr \nu^{}_\tau
}\right)^{}_{\rm L} = V \left( \matrix{\nu^{}_1 \cr \nu^{}_2 \cr
\nu^{}_3} \right)^{}_{\rm L} + R \left( \matrix{N^{}_1 \cr N^{}_2
\cr N^{}_3} \right)^{}_{\rm L} \; .
\end{eqnarray}
Then the standard charged current interactions between
$\nu^{}_\alpha$ and $\alpha$ (for $\alpha = e, \mu, \tau$) in the
basis of flavor states,
\begin{eqnarray}
-{\cal L}^{}_{\rm cc} \; = \; \frac{g}{\sqrt{2}} ~
\overline{\left(e~~ \mu~~ \tau\right)^{}_{\rm L}} ~ \gamma^\mu
\left( \matrix{\nu^{}_e \cr \nu^{}_\mu \cr \nu^{}_\tau}
\right)^{}_{\rm L} W^-_{\mu} ~ + ~ {\rm h.c.} \; ,
\end{eqnarray}
turn out to be of the form given in Eq. (1) in the basis of mass
states. It is clear that $V$ and $R$ describe the charged current
interactions of three light and heavy Majorana neutrinos,
respectively. While $V$ can be measured from a variety of neutrino
oscillation experiments, $R$ may be determined from possible
collider signatures of $N^{}_i$.

Note that $V$ itself is not unitary. Indeed, $V V^\dagger + R
R^\dagger = {\bf 1}$ holds as a consequence of unitarity of the
$6\times 6$ transformation matrix ${\cal U}$ in Eq. (3)
\cite{Chao}. The non-unitarity of the MNS matrix $V$ is an
intrinsic feature of the seesaw models, no matter whether they are
of type I or of type II. Since $V$ and $R$ are two $3\times 3$
sub-matrices of ${\cal U}$, their elements must be correlated with
each other. To find out the explicit correlation between $V$ and
$R$, we may parametrize ${\cal U}$ in terms of 15 rotation angles
and 15 phase angles \cite{FN1}. Then the common parameters
appearing in both $V$ and $R$ characterize their correlation.
First of all, let us define the 2-dimensional (1,2), (1,3) and
(2,3) rotation matrices in a 6-dimensional complex space:
\begin{eqnarray}
O^{}_{12} & = & \left( \matrix{c^{}_{12} & \hat{s}^*_{12} & 0 & 0
& 0 & 0 \cr -\hat{s}^{}_{12} & c^{}_{12} & 0 & 0 & 0 & 0 \cr 0 & 0
& 1 & 0 & 0 & 0 \cr 0 & 0 & 0 & 1 & 0 & 0 \cr 0 & 0 & 0 & 0 & 1 &
0 \cr 0 & 0 & 0 & 0 & 0 & 1 \cr} \right) \; , \nonumber \\
O^{}_{13} & = & \left( \matrix{c^{}_{13} & 0 & \hat{s}^*_{13} & 0
& 0 & 0 \cr 0 & 1 & 0 & 0 & 0 & 0 \cr -\hat{s}^{}_{13} & 0 &
c^{}_{13} & 0 & 0 & 0 \cr 0 & 0 & 0 & 1 & 0 & 0 \cr 0 & 0 & 0 & 0
& 1 & 0 \cr 0 & 0 & 0 & 0 & 0 & 1 \cr} \right) \; , \nonumber \\
O^{}_{23} & = & \left( \matrix{1 & 0 & 0 & 0 & 0 & 0 \cr 0 &
c^{}_{23} & \hat{s}^*_{23} & 0 & 0 & 0 \cr 0 & -\hat{s}^{}_{23} &
c^{}_{23} & 0 & 0 & 0 \cr 0 & 0 & 0 & 1 & 0 & 0 \cr 0 & 0 & 0 & 0
& 1 & 0 \cr 0 & 0 & 0 & 0 & 0 & 1 \cr} \right) \; ,
\end{eqnarray}
where $c^{}_{ij} \equiv \cos\theta^{}_{ij}$ and $\hat{s} \equiv
e^{i\delta^{}_{ij}} \sin\theta^{}_{ij}$ with $\theta^{}_{ij}$ and
$\delta^{}_{ij}$ being the rotation angle and phase angle,
respectively. Other 2-dimensional rotation matrices $O^{}_{ij}$
(for $1\leq i<j \leq 6$) can be defined in an analogous way
\cite{Fritzsch}. We parametrize the $6\times 6$ unitary matrix
${\cal U}$ as
\begin{equation}
{\cal U} \; = \; \left( \matrix{A & R \cr B & U \cr} \right)
\left( \matrix{V^{}_0 & {\bf 0} \cr {\bf 0} & {\bf 1} \cr} \right)
\end{equation}
with
\begin{eqnarray}
\left( \matrix{A & R \cr B & U \cr} \right) & = & O^{}_{56}
O^{}_{46} O^{}_{36} O^{}_{26} O^{}_{16} O^{}_{45} O^{}_{35}
O^{}_{25} O^{}_{15} O^{}_{34} O^{}_{24} O^{}_{14} \; , \nonumber
\\
\left( \matrix{V^{}_0 & {\bf 0} \cr {\bf 0} & {\bf 1} \cr} \right)
& = & O^{}_{23} O^{}_{13} O^{}_{12} \; .
\end{eqnarray}
Comparing between Eqs. (3) and (7), we get $V = A V^{}_0$ and $S =
B V^{}_0$, in which
\begin{equation}
V^{}_0 = \left( \matrix{ c^{}_{12} c^{}_{13} & \hat{s}^*_{12}
c^{}_{13} & \hat{s}^*_{13} \cr -\hat{s}^{}_{12} c^{}_{23} -
c^{}_{12} \hat{s}^{}_{13} \hat{s}^*_{23} & c^{}_{12} c^{}_{23} -
\hat{s}^*_{12} \hat{s}^{}_{13} \hat{s}^*_{23} & c^{}_{13}
\hat{s}^*_{23} \cr \hat{s}^{}_{12} \hat{s}^{}_{23} - c^{}_{12}
\hat{s}^{}_{13} c^{}_{23} & -c^{}_{12} \hat{s}^{}_{23} -
\hat{s}^*_{12} \hat{s}^{}_{13} c^{}_{23} & c^{}_{13} c^{}_{23}
\cr} \right)
\end{equation}
is just the standard parametrization of the unitary MNS matrix (up
to some proper phase rearrangements) \cite{PDG}. It is obvious
that $V \rightarrow V^{}_0$ in the limit of $A \rightarrow {\bf
1}$ (or equivalently, $R \rightarrow {\bf 0}$ and $S \rightarrow
{\bf 0}$). Thus $A$ signifies the unitarity violation of $V$.
After a lengthy but straightforward calculation, we obtain the
explicit expressions of $A$ and $R$ as follows:
\begin{eqnarray}
A & = & \left( \matrix{ c^{}_{14} c^{}_{15} c^{}_{16} & 0 & 0
\cr\cr
\begin{array}{l} -c^{}_{14} c^{}_{15} \hat{s}^{}_{16} \hat{s}^*_{26} -
c^{}_{14} \hat{s}^{}_{15} \hat{s}^*_{25} c^{}_{26} \\
-\hat{s}^{}_{14} \hat{s}^*_{24} c^{}_{25} c^{}_{26} \end{array} &
c^{}_{24} c^{}_{25} c^{}_{26} & 0 \cr\cr
\begin{array}{l} -c^{}_{14} c^{}_{15} \hat{s}^{}_{16} c^{}_{26} \hat{s}^*_{36}
+ c^{}_{14} \hat{s}^{}_{15} \hat{s}^*_{25} \hat{s}^{}_{26} \hat{s}^*_{36} \\
- c^{}_{14} \hat{s}^{}_{15} c^{}_{25} \hat{s}^*_{35} c^{}_{36} +
\hat{s}^{}_{14} \hat{s}^*_{24} c^{}_{25} \hat{s}^{}_{26}
\hat{s}^*_{36} \\
+ \hat{s}^{}_{14} \hat{s}^*_{24} \hat{s}^{}_{25} \hat{s}^*_{35}
c^{}_{36} - \hat{s}^{}_{14} c^{}_{24} \hat{s}^*_{34} c^{}_{35}
c^{}_{36} \end{array} &
\begin{array}{l} -c^{}_{24} c^{}_{25} \hat{s}^{}_{26} \hat{s}^*_{36} -
c^{}_{24} \hat{s}^{}_{25} \hat{s}^*_{35} c^{}_{36} \\
-\hat{s}^{}_{24} \hat{s}^*_{34} c^{}_{35} c^{}_{36} \end{array} &
c^{}_{34} c^{}_{35} c^{}_{36} \cr} \right) \; , \nonumber \\
R & = & \left( \matrix{ \hat{s}^*_{14} c^{}_{15} c^{}_{16} &
\hat{s}^*_{15} c^{}_{16} & \hat{s}^*_{16} \cr\cr
\begin{array}{l} -\hat{s}^*_{14} c^{}_{15} \hat{s}^{}_{16} \hat{s}^*_{26} -
\hat{s}^*_{14} \hat{s}^{}_{15} \hat{s}^*_{25} c^{}_{26} \\
+ c^{}_{14} \hat{s}^*_{24} c^{}_{25} c^{}_{26} \end{array} & -
\hat{s}^*_{15} \hat{s}^{}_{16} \hat{s}^*_{26} + c^{}_{15}
\hat{s}^*_{25} c^{}_{26} & c^{}_{16} \hat{s}^*_{26} \cr\cr
\begin{array}{l} -\hat{s}^*_{14} c^{}_{15} \hat{s}^{}_{16} c^{}_{26} \hat{s}^*_{36}
+ \hat{s}^*_{14} \hat{s}^{}_{15} \hat{s}^*_{25} \hat{s}^{}_{26} \hat{s}^*_{36} \\
- \hat{s}^*_{14} \hat{s}^{}_{15} c^{}_{25} \hat{s}^*_{35}
c^{}_{36} - c^{}_{14} \hat{s}^*_{24} c^{}_{25} \hat{s}^{}_{26}
\hat{s}^*_{36} \\
- c^{}_{14} \hat{s}^*_{24} \hat{s}^{}_{25} \hat{s}^*_{35}
c^{}_{36} + c^{}_{14} c^{}_{24} \hat{s}^*_{34} c^{}_{35} c^{}_{36}
\end{array} &
\begin{array}{l} -\hat{s}^*_{15} \hat{s}^{}_{16} c^{}_{26} \hat{s}^*_{36} -
c^{}_{15} \hat{s}^*_{25} \hat{s}^{}_{26} \hat{s}^*_{36} \\
+c^{}_{15} c^{}_{25} \hat{s}^*_{35} c^{}_{36} \end{array} &
c^{}_{16} c^{}_{26} \hat{s}^*_{36} \cr} \right) \; .
\end{eqnarray}
One can see that $A$ and $R$ involve the same parameters: nine
rotation angles ($\theta^{}_{i4}$, $\theta^{}_{i5}$ and
$\theta^{}_{i6}$ for $i=1$, $2$ and $3$) and nine phase angles
($\delta^{}_{i4}$, $\delta^{}_{i5}$ and $\delta^{}_{i6}$ for
$i=1$, $2$ and $3$). If all of them are switched off, we shall be
left with $R = {\bf 0}$ and $A = {\bf 1}$. Nontrivial $A$ is
therefore the bridge between $V = A V^{}_0$ and $R$.

Considering the fact that the unitarity violation of $V$ must be a
small effect (at most at the $1\%$ level as constrained by current
neutrino oscillation data and precision electroweak data
\cite{Antusch}), we may treat $A$ as a perturbation to $V^{}_0$.
The smallness of $\theta^{}_{ij}$ (for $i=1,2,3$ and $j=4,5,6$)
allows us to make the excellent approximations
\begin{eqnarray}
A & = & {\bf 1} - \left( \matrix{ \frac{1}{2} \left( s^2_{14} +
s^2_{15} + s^2_{16} \right) & 0 & 0 \cr \hat{s}^{}_{14}
\hat{s}^*_{24} + \hat{s}^{}_{15} \hat{s}^*_{25} + \hat{s}^{}_{16}
\hat{s}^*_{26} & \frac{1}{2} \left( s^2_{24} + s^2_{25} + s^2_{26}
\right) & 0 \cr \hat{s}^{}_{14} \hat{s}^*_{34} + \hat{s}^{}_{15}
\hat{s}^*_{35} + \hat{s}^{}_{16} \hat{s}^*_{36} & \hat{s}^{}_{24}
\hat{s}^*_{34} + \hat{s}^{}_{25} \hat{s}^*_{35} + \hat{s}^{}_{26}
\hat{s}^*_{36} & \frac{1}{2} \left( s^2_{34} +
s^2_{35} + s^2_{36} \right) \cr} \right) + {\cal O}(s^4_{ij}) \; , \nonumber \\
R & = & {\bf 0} + \left( \matrix{ \hat{s}^*_{14} & \hat{s}^*_{15}
& \hat{s}^*_{16} \cr \hat{s}^*_{24} & \hat{s}^*_{25} &
\hat{s}^*_{26} \cr \hat{s}^*_{34} & \hat{s}^*_{35} &
\hat{s}^*_{36} \cr} \right) + {\cal O}(s^3_{ij}) \; ,
\end{eqnarray}
from which one can easily check the validity of $VV^\dagger =
AA^\dagger = {\bf 1} - RR^\dagger$ to a good degree of accuracy.
Thus the nine mixing angles in Eq. (10) or (11) are all of ${\cal
O}(10^{-1})$ or smaller, such that the unitarity violation of $V$
can maximally be of ${\cal O}(10^{-2})$. The nine CP-violating
phases of $A$ or $R$ are in general not suppressed, however. It is
worth remarking that $R \sim {\cal O}(10^{-3})$ to ${\cal
O}(10^{-1})$ may lead to appreciable collider signatures of lepton
number violation, if the masses of heavy Majorana neutrinos $M^{}_i$
are of ${\cal O}(10^2)$ GeV to ${\cal O}(10)$ TeV. At the LHC, for
instance, the promising lepton-number-violating processes mediated
by $N^{}_i$ include $pp \to W^\pm W^\pm \to \mu^\pm \mu^\pm jj$ and
$pp \to W^\pm \to \mu^\pm N \to \mu^\pm \mu^\pm jj$, where the
$\Delta L =2$ like-sign dilepton production can unambiguously signal
the existence of $N^{}_i$ \cite{Han,LHC,Smirnov}. A preliminary
analysis made in Ref. \cite{Han} has shown that it is possible to
probe $(RR^\dagger)^{}_{\mu\mu} \approx s^2_{24} + s^2_{25} +
s^2_{26}$ to ${\cal O}(10^{-4})$ for $M^{}_i \sim 100$ GeV and to
${\cal O}(10^{-2})$ for $M^{}_i \sim 400$ GeV at the LHC with an
integrated luminosity $100 ~{\rm fb^{-1}}$. The sensitivity will in
general become worse for much larger values of the heavy Majorana
neutrino masses.

Note that the triangular form of $A$ is a salient feature of our
parametrization. Some straightforward consequences on $V =
AV^{}_0$ can be obtained from Eqs. (9) and (10).
\begin{itemize}
\item     $V^{}_{e3} = c^{}_{14} c^{}_{15} c^{}_{16} \hat{s}^*_{13}$
holds. Given $\theta^{}_{13} = 0$ for $V^{}_0$, which might result
from certain flavor symmetries imposed on $M^{}_{\rm L}$,
$M^{}_{\rm D}$ and $M^{}_{\rm R}$ \cite{Chao,Smirnov}, $V^{}_{e3}$
turns out to vanish.

\item     The ratio $|V^{}_{e2}/V^{}_{e1}| = \tan\theta^{}_{12}$
is completely irrelevant to the parameters appearing in $A$ or
$R$. This result implies that the extraction of $\theta^{}_{12}$
from the solar neutrino oscillation data is essentially
independent of possible unitarity violation of $V$.

\item     $|V^{}_{e1}|^2 + |V^{}_{e2}|^2 + |V^{}_{e3}|^2 =
c^2_{14} c^2_{15} c^2_{16} \approx 1 - (s^2_{14} + s^2_{15} +
s^2_{16})$ holds. Hence non-vanishing $\theta^{}_{14}$,
$\theta^{}_{15}$ and $\theta^{}_{16}$ violate the normalization
condition of three matrix elements in the first row of $V$.
Current experimental data require $(s^2_{14} + s^2_{15} +
s^2_{16}) \lesssim 1\%$ \cite{Antusch}.

\item     $\langle m\rangle^{}_{ee} = c^2_{14} c^2_{15} c^2_{16}
|m^{}_1 (c^{}_{12} c^{}_{13})^2 + m^{}_2 (\hat{s}^*_{12}
c^{}_{13})^2 + m^{}_3 (\hat{s}^*_{13})^2|$ holds for the effective
mass of the neutrinoless double-beta decay. The smallness of
$\theta^{}_{14}$, $\theta^{}_{15}$ and $\theta^{}_{16}$ implies
that their effects on $\langle m\rangle^{}_{ee}$ are in practice
negligible.

\item     $\langle m\rangle^{}_{e} = c^{}_{14} c^{}_{15} c^{}_{16}
\sqrt{m^2_1 c^2_{12} c^2_{13} + m^2_2 s^2_{12} c^2_{13} + m^2_3
s^2_{13}}$ holds for the effective mass term of the tritium beta
decay. The smallness of $\theta^{}_{14}$, $\theta^{}_{15}$ and
$\theta^{}_{16}$ implies that their effects on $\langle
m\rangle^{}_{e}$ are also negligible.
\end{itemize}
Another consequence of the non-unitarity of $V$ is the loss of
universality for the Jarlskog invariants of CP violation \cite{J},
$J^{ij}_{\alpha\beta} \equiv {\rm Im}(V^{}_{\alpha i} V^{}_{\beta
j} V^*_{\alpha j} V^*_{\beta i})$, where the Greek indices run
over $(e, \mu, \tau)$ and the Latin indices run over $(1,2,3$).
The explicit expressions of $J^{ij}_{\alpha\beta}$ in terms of
$\theta^{}_{ij}$ and $\delta^{}_{ij}$ are rather complicated and
will be presented elsewhere. But we shall illustrate that the
extra CP-violating phases of $V$ are possible to give rise to a
significant asymmetry between $\nu^{}_\mu \rightarrow \nu^{}_\tau$
and $\overline{\nu}^{}_\mu \rightarrow \overline{\nu}^{}_\tau$
oscillations with short or medium baselines in the subsequent
section.

\vspace{0.4cm}

\framebox{\large\bf 3} ~ For the sake of simplicity, here we only
consider a special but interesting pattern of $V^{}_0$:
$\tan\theta^{}_{12} = 1/\sqrt{2}$, $\theta^{}_{13} = 0$ and
$\theta^{}_{23} =\pi/4$ (namely, $V^{}_0$ takes the well-known
tri-bimaximal mixing pattern \cite{TB}). Then the non-unitary
neutrino mixing matrix $V = A V^{}_0$ can approximate to
\begin{equation}
V \; \approx \; \left( \matrix{ \sqrt{\frac{2}{3}} &
\sqrt{\frac{1}{3}} ~ e^{-i\delta^{}_{12}} & 0 \cr
-\sqrt{\frac{1}{6}} ~\left( 1 + 2 X \right) e^{i\delta^{}_{12}} &
\sqrt{\frac{1}{3}} ~\left( 1 - X \right) & \sqrt{\frac{1}{2}} ~
e^{-i\delta^{}_{23}} \cr \sqrt{\frac{1}{6}} ~\left( 1 - 2Y + Z
\right) e^{i (\delta^{}_{12} + \delta^{}_{23})} &
-\sqrt{\frac{1}{3}} ~\left( 1 + Y + Z \right) e^{i\delta^{}_{23}}
& \sqrt{\frac{1}{2}} ~\left( 1 - Z \right) \cr} \right) \; ,
\end{equation}
where
\begin{eqnarray}
X & = & \left( \hat{s}^{}_{14} \hat{s}^*_{24} + \hat{s}^{}_{15}
\hat{s}^*_{25} + \hat{s}^{}_{16} \hat{s}^*_{26} \right)
e^{-i\delta^{}_{12}} \; , \nonumber \\
Y & = & \left( \hat{s}^{}_{14} \hat{s}^*_{34} + \hat{s}^{}_{15}
\hat{s}^*_{35} + \hat{s}^{}_{16} \hat{s}^*_{36} \right)
e^{-i(\delta^{}_{12} + \delta^{}_{23})} \; , \nonumber \\
Z & = & \left( \hat{s}^{}_{24} \hat{s}^*_{34} + \hat{s}^{}_{25}
\hat{s}^*_{35} + \hat{s}^{}_{26} \hat{s}^*_{36} \right)
e^{-i\delta^{}_{23}} \; .
\end{eqnarray}
The Jarlskog invariants $J^{ij}_{\alpha\beta}$ turn out to be
$J^{23}_{e\mu} = J^{23}_{\tau e} = J^{31}_{e\mu} = J^{31}_{\tau e} =
0$, $J^{12}_{e\mu} \approx {\rm Im}X/3$, $J^{12}_{\tau e} \approx
{\rm Im}Y/3$, and
\begin{eqnarray}
J^{12}_{\mu\tau} & \approx & \left( {\rm Im}X + {\rm Im}Y
\right)/6 \; , \nonumber \\
J^{23}_{\mu\tau} & \approx & \left( {\rm Im}X + {\rm
Im}Y + 2{\rm Im}Z \right)/6 \; , \nonumber \\
J^{31}_{\mu\tau} & \approx & \left( {\rm Im}X + {\rm Im}Y - {\rm
Im}Z \right)/6 \; .
\end{eqnarray}
Note that $J^{ji}_{\alpha\beta} = J^{ij}_{\beta\alpha} =
-J^{ij}_{\alpha\beta}$ holds as a direct consequence of the
definition of $J^{ij}_{\alpha\beta}$. It becomes clear that
different $J^{ij}_{\alpha\beta}$ may in general have different
values in the presence of unitarity violation, which can result in
some new CP-violating effects in neutrino oscillations via the
phase parameters $\delta^{}_{kl}$ (for $k=1,2,3$ and $l=4,5,6$)
hidden in $X$, $Y$ and $Z$.

Taking the steps outlined in Ref. \cite{Antusch}, one may easily
derive the probabilities of $\nu^{}_\alpha \rightarrow
\nu^{}_\beta$ oscillations in vacuum. The result is
\footnote{Note that the signs of three CP-violating terms in our
Eq. (15) are opposite to those in Eq. (18) of Ref. \cite{Antusch}.
The latter might result from a misprint.}
\begin{equation}
P(\nu^{}_\alpha \rightarrow \nu^{}_\beta) \; = \;
\frac{\displaystyle \sum^{}_i |V^{}_{\alpha i}|^2 |V^{}_{\beta
i}|^2 + 2 \sum^{}_{i<j} {\rm Re} \left( V^{}_{\alpha i}
V^{}_{\beta j} V^*_{\alpha j} V^*_{\beta i} \right) \cos
\Delta^{}_{ij} - 2 \sum^{}_{i<j} J^{ij}_{\alpha\beta}
\sin\Delta^{}_{ij}}{\displaystyle \left(
VV^\dagger\right)^{}_{\alpha\alpha} \left(
VV^\dagger\right)^{}_{\beta\beta}} \; ,
\end{equation}
where $\Delta^{}_{ij} \equiv \Delta m^2_{ij} L/(2E)$ with $\Delta
m^2_{ij} \equiv m^2_i - m^2_j$, $E$ being the neutrino beam energy
and $L$ being the baseline length. It is straightforward to write
out the expression of $P(\overline{\nu}^{}_\alpha \rightarrow
\overline{\nu}^{}_\beta)$ from Eq. (15) by making the replacement
$V \Longrightarrow V^*$ or equivalently $J^{ij}_{\alpha\beta}
\Longrightarrow - J^{ij}_{\alpha\beta}$. If $V$ is exactly unitary
(i.e., $A = {\bf 1}$ and $V = V^{}_0$), the denominator of Eq.
(15) will become unity and the conventional formula of
$P(\nu^{}_\alpha \rightarrow \nu^{}_\beta)$ will be reproduced. It
has been observed in Ref. \cite{Yasuda} that $\nu^{}_\mu
\rightarrow \nu^{}_\tau$ and $\overline{\nu}^{}_\mu \rightarrow
\overline{\nu}^{}_\tau$ oscillations may serve as an excellent
tool to probe possible signatures of CP violation induced by the
non-unitarity of $V$. To see this point more clearly, we consider
a short- or medium-baseline neutrino oscillation experiment with
$|\sin\Delta^{}_{13}| \sim |\sin\Delta^{}_{23}| \gg
|\sin\Delta^{}_{12}|$, in which the terrestrial matter effects are
expected to be insignificant or negligibly small. Then the
dominant CP-conserving and CP-violating terms of $P(\nu^{}_\mu
\rightarrow \nu^{}_\tau)$ and $P(\overline{\nu}^{}_\mu \rightarrow
\overline{\nu}^{}_\tau)$ can simply be obtained from Eq. (15)
\footnote{Note that the CP-violating term shown in our Eq. (16) is
apparently different from that given in Eq. (12) of Ref.
\cite{Yasuda}, where a very different parametrization of the
unitarity violation of $V$ has been adopted.}:
\begin{eqnarray}
P(\nu^{}_\mu \rightarrow \nu^{}_\tau) & \approx & \sin^2
2\theta^{}_{23} \sin^2 \frac{\Delta^{}_{23}}{2} ~ - ~ 2 \left(
J^{23}_{\mu\tau} + J^{13}_{\mu\tau} \right) \sin\Delta^{}_{23} \; ,
\nonumber \\
P(\overline{\nu}^{}_\mu \rightarrow \overline{\nu}^{}_\tau) &
\approx & \sin^2 2\theta^{}_{23} \sin^2 \frac{\Delta^{}_{23}}{2} ~
+ ~ 2 \left( J^{23}_{\mu\tau} + J^{13}_{\mu\tau} \right)
\sin\Delta^{}_{23} \; ,
\end{eqnarray}
where the good approximation $\Delta^{}_{13} \approx
\Delta^{}_{23}$ has been used in view of the experimental fact
$|\Delta m^2_{13}| \approx |\Delta m^2_{23}| \gg |\Delta
m^2_{12}|$ \cite{SNO,SK,KM,K2K}, and the sub-leading and
CP-conserving ``zero-distance" effect \cite{Antusch} has been
omitted. Taking account of Eq. (14), we find $J^{23}_{\mu\tau} +
J^{13}_{\mu\tau} = J^{23}_{\mu\tau} - J^{31}_{\mu\tau} \approx
{\rm Im}Z/2$. To be more specific,
\begin{eqnarray}
J^{23}_{\mu\tau} + J^{13}_{\mu\tau} & \approx & \frac{1}{2} \left[
s^{}_{24} s^{}_{34} \sin \left( \delta^{}_{24} - \delta^{}_{34} -
\delta^{}_{23} \right) ~ + ~ s^{}_{25} s^{}_{35} \sin \left(
\delta^{}_{25} - \delta^{}_{35} - \delta^{}_{23} \right) \right .
\nonumber \\
&& \left . + ~ s^{}_{26} s^{}_{36} \sin \left( \delta^{}_{26} -
\delta^{}_{36} - \delta^{}_{23} \right) \right] \; .
\end{eqnarray}
In the assumption of $s^{}_{2l} \sim s^{}_{3l} \sim 0.1$ and
$(\delta^{}_{2l} - \delta^{}_{3l}) \sim 1$ (for $l=4,5,6$), this
non-trivial CP-violating quantity can reach the percent level.

FIG. 1 illustrates the CP-violating asymmetry between $\nu^{}_\mu
\rightarrow \nu^{}_\tau$ and $\overline{\nu}^{}_\mu \rightarrow
\overline{\nu}^{}_\tau$ oscillations, where $\theta^{}_{23} \sim
\pi/4$, $\Delta m^2_{23} \sim 2.5 \times 10^{-3} ~ {\rm eV}^2$ and
$(J^{23}_{\mu\tau} + J^{13}_{\mu\tau}) \sim 0.01$ have typically
been input. One can see that it is possible to measure this
asymmetry in the range $L/E \sim (100 \cdots 400)$ km/GeV, if the
experimental sensitivity is $\lesssim 1\%$. A short- or
medium-baseline neutrino factory with the beam energy $E$ being
above $m^{}_\tau \approx 1.78$ GeV is expected to have a good chance
to probe or constrain the effect of CP violation induced by the
non-unitarity of $V$ (see Ref. \cite{Yasuda} for some detailed
discussions).

\vspace{0.4cm}

\framebox{\large\bf 4} ~ No matter whether a neutrino mass model
is of Type-I seesaw or Type-II seesaw, nine rotation angles and
nine phase angles are in general needed to parametrize the
correlation between the charged current interactions of three
heavy Majorana neutrinos $(N^{}_1, N^{}_2, N^{}_3)$ and their
light counterparts $(\nu^{}_1, \nu^{}_2, \nu^{}_3)$. To reduce the
number of free parameters, one may consider the ``unbalanced"
seesaw scenarios in which the number of heavy Majorana neutrinos
is smaller than that of ligh Majorana neutrinos \cite{Xing07}.
There are at least two phenomenologically viable scenarios of this
nature: one is the minimal Type-I seesaw model with two heavy
Majorana neutrinos \cite{FGY} and the other is the minimal Type-II
seesaw model with one heavy Majorana neutrino and one Higgs
triplet \cite{Chan}. In either case, we can arrive at the
simplified correlation between $V$ and $R$.

(1) The minimal Type-I seesaw model with two heavy Majorana
neutrinos. In this case, $M^{}_{\rm R}$ is $2\times 2$, $M^{}_{\rm
D}$ is $3\times 2$, and $M^{}_{\rm L} = {\bf 0}$ holds in the
overall $5\times 5$ neutrino mass matrix ${\cal M}$. Switching off
the rotation matrices $O^{}_{i6}$ (for $i=1,\cdots,5$) in Eq. (8),
we are able to fully parametrize the $3\times 3$ matrix $A$ and
the $3\times 2$ matrix $R$ in terms of six rotation angles
($\theta^{}_{i4}$ and $\theta^{}_{i5}$ for $i=1,2,3$) and six
phase angles ($\delta^{}_{i4}$ and $\delta^{}_{i5}$ for
$i=1,2,3$):
\begin{eqnarray}
A & = & \left( \matrix{ c^{}_{14} c^{}_{15} & 0 & 0 \cr -c^{}_{14}
\hat{s}^{}_{15} \hat{s}^*_{25} -\hat{s}^{}_{14} \hat{s}^*_{24}
c^{}_{25} & c^{}_{24} c^{}_{25} & 0 \cr -c^{}_{14} \hat{s}^{}_{15}
c^{}_{25} \hat{s}^*_{35} + \hat{s}^{}_{14} \hat{s}^*_{24}
\hat{s}^{}_{25} \hat{s}^*_{35} - \hat{s}^{}_{14} c^{}_{24}
\hat{s}^*_{34} c^{}_{35} & -c^{}_{24} \hat{s}^{}_{25}
\hat{s}^*_{35} - \hat{s}^{}_{24} \hat{s}^*_{34}
c^{}_{35} & c^{}_{34} c^{}_{35} \cr} \right) \; , ~~~~~~ \nonumber \\
R & = & \left( \matrix{ \hat{s}^*_{14} c^{}_{15} & \hat{s}^*_{15}
\cr -\hat{s}^*_{14} \hat{s}^{}_{15} \hat{s}^*_{25} + c^{}_{14}
\hat{s}^*_{24} c^{}_{25} & c^{}_{15} \hat{s}^*_{25} \cr
-\hat{s}^*_{14} \hat{s}^{}_{15} c^{}_{25} \hat{s}^*_{35} -
c^{}_{14} \hat{s}^*_{24} \hat{s}^{}_{25} \hat{s}^*_{35} +
c^{}_{14} c^{}_{24} \hat{s}^*_{34} c^{}_{35} & c^{}_{15} c^{}_{25}
\hat{s}^*_{35} \cr} \right) \; .
\end{eqnarray}
It is therefore straightforward to obtain the correlation between
$V = AV^{}_0$ and $R$.

(2) The minimal Type-II seesaw model with one heavy Majorana
neutrino and one Higgs triplet. In this case, $M^{}_{\rm R}$ is
$1\times 1$, $M^{}_{\rm D}$ is $3\times 1$, and $M^{}_{\rm L}$ is
$3\times 3$ in the overall $4\times 4$ neutrino mass matrix ${\cal
M}$. Switching off the rotation matrices $O^{}_{i5}$ (for $i < 5$)
and $O^{}_{i6}$ (for $i < 6$) in Eq. (8), we can parametrize the
$3\times 3$ matrix $A$ and the $3\times 1$ matrix $R$ in terms of
three rotation angles ($\theta^{}_{i4}$ for $i=1,2,3$) and three
phase angles ($\delta^{}_{i4}$ for $i=1,2,3$) as follows:
\begin{eqnarray}
A & = & \left( \matrix{ c^{}_{14} & 0 & 0 \cr -\hat{s}^{}_{14}
\hat{s}^*_{24} & c^{}_{24} & 0 \cr - \hat{s}^{}_{14} c^{}_{24}
\hat{s}^*_{34} & - \hat{s}^{}_{24} \hat{s}^*_{34} & c^{}_{34} \cr}
\right) \; , \nonumber \\
R & = & \left( \matrix{ \hat{s}^*_{14} \cr c^{}_{14}
\hat{s}^*_{24} \cr c^{}_{14} c^{}_{24} \hat{s}^*_{34} \cr} \right)
\; .
\end{eqnarray}
Of course, the correlation between $V = AV^{}_0$ and $R$ is more
obvious in this scenario. Taking the Jarlskog invariant
$J^{23}_{e\mu}$ for example, we find
\begin{eqnarray}
J^{23}_{e\mu} & = & s^{}_{12} c^{}_{13} s^{}_{13} c^2_{14}
c^{}_{24} \left[ c^{}_{12} c^{}_{13} c^{}_{23} s^{}_{23} c^{}_{24}
\sin \left( \delta^{}_{13} - \delta^{}_{12} - \delta^{}_{23}
\right) \right . \nonumber \\
&& \left . - c^{}_{12} s^{}_{13} s^{}_{14} c^{}_{23} s^{}_{24}
\sin \left( \delta^{}_{14} - \delta^{}_{12} - \delta^{}_{24}
\right) + s^{}_{12} s^{}_{14} s^{}_{23} s^{}_{24} \sin \left(
\delta^{}_{14} - \delta^{}_{13} + \delta^{}_{23} - \delta^{}_{24}
\right) \right] \; ,
\end{eqnarray}
in which the first term is essentially governed by the phase
parameters of $V^{}_0$, but the second and third terms result from
the unitarity-violating effects. One may similarly calculate the
other Jarlskog invariants of CP violation. As for $\nu^{}_\mu
\rightarrow \nu^{}_\tau$ and $\overline{\nu}^{}_\mu \rightarrow
\overline{\nu}^{}_\tau$ oscillations with short or medium
baselines, the approximate probabilities obtained in Eq. (16) are
generally applicable and the CP-violating quantity
$(J^{23}_{\mu\tau} + J^{13}_{\mu\tau})$ may unambiguously signify
the non-unitarity of $V$ or the existence of certain non-standard
neutrino interactions.

\vspace{0.4cm}

\framebox{\large\bf 5} ~ In summary, we have investigated how the
charged current interactions of light and heavy Majorana neutrinos
are correlated with each other in the Type-I and Type-II seesaw
models. It is the first time that an explicit parametrization of
this correlation, which is independent of any details of the
seesaw models, has been presented to bridge the gap between
neutrino physics and collider physics. The rotation angles and
phase angles in such a parametrization are expected to be measured
or constrained in the precision neutrino oscillation experiments
and by exploring possible signatures of heavy Majorana neutrinos
at the LHC and ILC. We have taken two special but viable examples,
the minimal Type-I seesaw model with two heavy Majorana neutrinos
and the minimal Type-II seesaw model with one heavy Majorana
neutrino and one Higgs triplet, to illustrate the simplified
$V$-$R$ correlation. The implications of $R\neq 0$ on the
low-energy neutrino phenomenology, such as the neutrinoless
double-beta decay, the tritium beta decay and CP violation in
neutrino oscillations, have also been discussed. In particular, we
have demonstrated that the non-unitarity of $V$ is possible to
give rise to an appreciable CP-violating asymmetry between
$\nu^{}_\mu \rightarrow \nu^{}_\tau$ and $\overline{\nu}^{}_\mu
\rightarrow \overline{\nu}^{}_\tau$ oscillations with short or
medium baselines. Our generic results remain valid even if the
Type-I or Type-II seesaw mechanism is embedded in the
supersymmetric standard model and some other extensions of the
standard model.

How to naturally realize an appreciable correlation between $V$
and $R$ in a TeV-scale seesaw model is actually a real challenge
to model builders. The reason is simply that the main textures of
$M^{}_{\rm D}$ and $M^{}_{\rm R}$ in the Type-I seesaw scenarios
or those of $M^{}_{\rm D}$, $M^{}_{\rm L}$ and $M^{}_{\rm R}$ in
the Type-II seesaw scenarios, which are relevant to possibly
observable collider signatures, are difficult to imprint on those
sub-leading effects (due to explicit perturbations or radiative
corrections) responsible for the tiny masses of light Majorana
neutrinos \cite{Chao,Smirnov}. Much more efforts are certainly
needed to build viable and natural seesaw models at the TeV scale.
One may even speculate the possibility to naturally achieve the
TeV-scale leptogenesis and to experimentally test it at the LHC
and ILC.

Let us stress that testing the unitarity of the light Majorana
neutrino mixing matrix in neutrino oscillations and searching for
the signatures of heavy Majorana neutrinos at TeV-scale colliders
can be complementary to each other, both qualitatively and
quantitatively, in order to deeply understand the intrinsic
properties of Majorana particles. Any experimental breakthrough in
this aspect will pave the way towards the true theory of neutrino
mass generation and flavor mixing.

\vspace{0.6cm}

The author would like to thank S. Zhou for useful discussions.
This work was supported in part by the National Natural Science
Foundation of China.

\newpage

\begin{figure}[t]
\vspace{0cm}
\epsfig{file=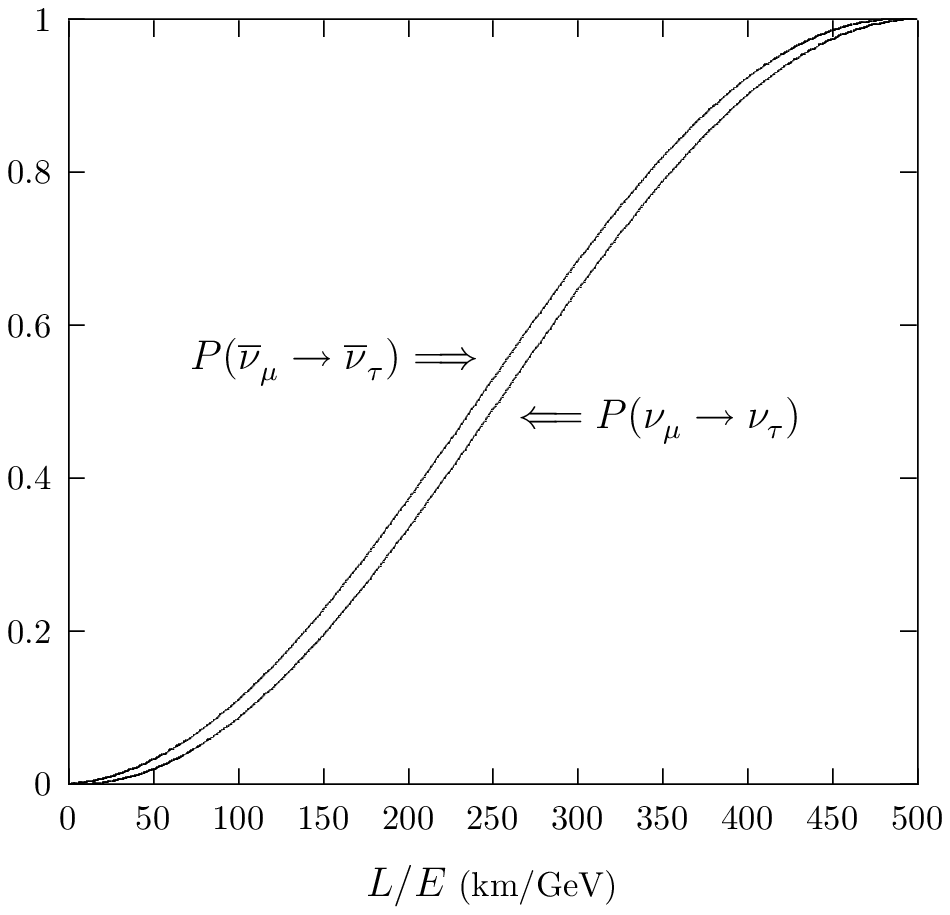,bbllx=2.5cm,bblly=14cm,bburx=12.3cm,bbury=29cm,%
width=10cm,height=16cm,angle=0,clip=0} \vspace{-0.9cm}
\caption{Illustration of the CP-violating asymmetry, which is
induced by the non-unitarity of $V$, between $\nu^{}_\mu
\rightarrow \nu^{}_\tau$ and $\overline{\nu}^{}_\mu \rightarrow
\overline{\nu}^{}_\tau$ oscillations with short or medium
baselines. Here $\theta^{}_{23} \sim \pi/4$, $\Delta m^2_{23} \sim
2.5 \times 10^{-3} ~ {\rm eV}^2$ and $(J^{23}_{\mu\tau} +
J^{13}_{\mu\tau}) \sim 0.01$ have typically been input.}
\end{figure}

\end{document}